\documentstyle[preprint,prc,aps,floats,epsf]{revtex}

\tightenlines

\newcommand{\beq}{\begin{equation}}
\newcommand{\eeq}{\end{equation}}
\newcommand{\be}{\begin{eqnarray}}
\newcommand{\ee}{\end{eqnarray}}
\newcommand{\ben}{\begin{eqnarray*}}
\newcommand{\een}{\end{eqnarray*}}
\newcommand{\drms}{DR[$\overline{\mbox{MS}}$]}

\begin{document}
\draft

\title{Regularization Methods for Nucleon-Nucleon\\ Effective Field Theory}

\author{James V. Steele and R. J. Furnstahl}
\address{Department of Physics \\
         The Ohio State University,\ \ Columbus, OH\ \ 43210}

\date{February, 1998}

\maketitle

\begin{abstract}
Attempts to apply effective field theory (EFT) methods
to nonrelativistic
nucleon-nucleon ($NN$) scattering have raised questions about
the nature and limitations 
of an EFT expansion when used nonperturbatively.
We discuss the characteristics of a meaningful EFT
analysis and compare them with traditional  approaches
to $NN$ scattering.  A key feature of an EFT treatment is a
systematic expansion in powers of momentum, which we demonstrate using 
an error analysis introduced by Lepage.  A clear graphical
determination of the radius of convergence for the momentum expansion
is also obtained.  We use these techniques to compare cutoff
regularization, two forms of dimensional regularization, and the  
dibaryon approach, using a simple model for illustration.   The
naturalness of the parameters and predictions for 
bound-state energies are also shown.
\end{abstract}


\thispagestyle{empty}

\newpage


\section{Introduction}

The study of nuclear forces and nucleon-nucleon ($NN$) scattering has
a long history~\cite{Review}.  Numerous models have evolved that provide
superb reproductions of data over large energy ranges
with only a few parameters per scattering channel.
Nevertheless, the models do not provide reliable error estimates, 
and their connection 
to the symmetries and dynamics of the underlying theory of the
strong interaction, quantum chromodynamics (QCD), is obscure.

Recently, many papers have addressed the application of
effective field theory (EFT) methods to $NN$ scattering 
\cite{BiraThesis,vanKolck,Weinberg,%
Maryland,Lepage,KSW1,LukeManohar,Kaplan,Bedaq,Richardson,KSW2,park,gege}.
The hope is that such an analysis will yield a systematic, controlled
expansion and bring new understanding to how
various processes involving nucleons relate to each other \cite{vanKolck}.
This requires a nonperturbative treatment of the EFT to properly
account for the interaction between two or more heavy
nucleons \cite{Weinberg}.  

There are currently many disagreements in the literature
about the nature and limitations of an EFT expansion in this 
case \cite{Maryland,Lepage,KSW1,LukeManohar,Kaplan,Richardson}.
Regularization is required to handle divergences that arise, but
the results are said to depend on 
the regularization scheme used (see
Table~\ref{tab1}) and 
the size of the scattering length involved.  
More generally, it is claimed that the behavior and
predictive power expected from a true effective field theory is
not exhibited by every
regularization method when applied nonperturbatively \cite{Maryland,Lepage}.
However, others claim that results are independent of the regularization
method when proper power counting is applied \cite{vanKolck,Bedaq}.

Our goal here is to clarify the important features of a
nonperturbative EFT.  We can do this most clearly by making a
side-by-side comparison of the regularization schemes in
Table~\ref{tab1} using the error analysis advocated by Lepage
\cite{Lepage}.  This comparison clearly illustrates which
schemes behave like a true EFT.
Those familiar with the successful phenomenological
models for $NN$ scattering such as the Reid \cite{Reid}, 
Bonn \cite{Review}, 
and Paris \cite{Review} potentials, which already reproduce
the data well, may wonder why there is a need to reformulate the
problem in terms of an EFT.  We use the Reid potential
as an example to compare  conventional
and EFT
approaches. 

\begin{table}[b] \caption{\label{tab1} Regularization schemes
and their abbreviations used throughout this paper.}
\begin{tabular}[tbh]{cp{5.5in}}
name & \hfill regularization scheme \hfill\null \\ \hline
CR[G] & Cutoff regularization with gaussian weighting in the potential
\cite{Lepage}
\\
\drms{} & Dimensional regularization with modified minimal subtraction
\cite{KSW1}
\\
DR[PDS] & Dimensional regularization with power divergence subtraction
\cite{KSW2}
\\
dibaryon & Additional low-energy degree of freedom associated with
         a bound or nearly bound state \cite{Kaplan,Bedaq}.
\end{tabular}
\end{table}

We start by reviewing the expected behavior and radius of convergence
of an EFT and outline the 
various regularization schemes in Section~II.  We then reiterate
Lepage's discussion on how to analyze  
effective field theory behavior in terms of error plots
\cite{Lepage} in Section~III. We also show how to numerically extend his
approach beyond second order in the momentum expansion and how to
determine the radius of convergence 
graphically from the error plots.  This procedure allows us to 
compare different regularization methods on the same footing.

Our experience with a variety of different models suggests that
the issues mentioned above are generic.
Therefore we use the simple delta-shell
potential without pion contributions as a convenient ``laboratory''
for comparing regularization methods in Section~IV.   
We find that all of the regularization
schemes considered in this paper have a proper radius of convergence
except for dimensional regularization with modified minimal
subtraction (\drms{}) when applied as in Ref.~\cite{KSW1}.
Furthermore,
a good radius of convergence is linked to the naturalness of the
constants from the effective lagrangian and the ability to make
reliable error estimates.  

An effective field theory should provide a low-energy realization
of the S-matrix.
This implies that 
after fitting the potential to a given order by the (scattering) phase
shifts, estimates of the bound state energies should have similar
error scaling.  We present results for the bound state 
in Section~IV and
our conclusions in Section~V.

\section{Defining an Effective Field Theory}

The strong interaction at low energy can be described by an effective
lagrangian of hadrons, whose form is constrained by the
symmetries of the underlying theory, QCD.  Restricting the
characteristic momentum $p$ of the interaction to be less than a scale
$\Lambda$ implies that physics at larger mass scales, such as from the
exchange of heavier particles, is not resolved.  This separates the
physics into a long-distance part, given by the light dynamical fields in
the effective lagrangian, and a short-distance part consisting of the
heavy degrees of freedom and entering only 
through renormalization of effective lagrangian couplings.
An exception is the nucleon,
which due to baryon number conservation can be
present as a heavy source.  

The pion is the lightest of the hadrons and a prime candidate to be
included as a dynamical
degree of freedom.  Its Goldstone nature leads to a systematic
treatment,
known as chiral perturbation theory \cite{ChPT},  
that has been thoroughly studied. 
No systematic treatment is known for non-Goldstone
particles such as the $\rho$ meson.  Its exclusion from  
the effective lagrangian sets the scale of new, underlying
physics $\Lambda\sim m_\rho$.  

We stress that the effects of the short-distance
physics on the long-distance
(low-momentum) observables can be reproduced by generic terms in the effective
lagrangian, organized in powers of
derivatives over $\Lambda$, or a momentum expansion \cite{Weinberg1}.  
The
coefficients of these terms summarize the remnants of the short-distance
physics.
In the future, these renormalized coefficients could in
principle be calculated on the lattice.  Until then, they must be
fixed by existing data.  

As each additional coefficient in the lagrangian is fixed, the error
in the S-matrix should improve by a power of 
$p^2/\Lambda^2$ \cite{Lepage}.  (If the interaction contains single
powers of momentum coupled, for example, to an external field, the
improvement may be by powers of $p/\Lambda$.)
This identification of the accuracy with
powers of $\Lambda$, called power counting, implies that
the calculation of every observable from the same EFT will 
give the same systematic momentum dependence on the error, as long as
the corresponding effective operator for each observable is determined
to the same order in $\Lambda$. 
Furthermore, the
dimensionless coefficients 
of the terms in the effective lagrangian are typically
of order unity (natural)
when the cutoff is chosen on the order of the physics not contained in
the lagrangian.  All the above features are favorable for
systematic predictions and an assessment of the corrections.

At best, this expansion breaks down when the momenta in the processes
involved are comparable to $\Lambda$, at which point the short-distance
physics begins to be resolved.  
This point is referred to as the radius of convergence
of the EFT.  
We will
see below ({\it e.g.\/}, the second plot of Fig.~\ref{phaseshift}) that
this also has a clear graphical interpretation as a convergence of the
error involved when an increasing number of orders in $p/\Lambda$ are
taken into account in the effective field theory.

Systems consisting of two (or more) nucleons with momenta 
well below the scale $\Lambda$ can be treated
nonrelativistically.  Interactions between the nucleons 
lead to infrared divergences that require a summation to all orders of a
certain class of diagrams \cite{Weinberg}.  
This can be achieved by solving the Schr\"odinger equation or,
equivalently, the Lippmann-Schwinger equation 
\be
   T = V + V G_0 T \ .
\label{LS}
\ee 
The short-distance
effective lagrangian is schematically mapped to a 
coordinate-space potential
consisting of highly singular terms with delta functions and derivatives
of delta functions.  Only after specifying a regularization
scheme is the effective potential $V$ defined. The energy dependence
of $V$ can be traded for  three-momentum dependence by using the
equations of motion \cite{Georgi}.

The different ways the potential can be regularized lead to different
candidate  
effective field theories.  
In perturbative applications the predictions of an EFT
are independent of the regularization method.  
However, this result has not been established for nonperturbative
applications; indeed, some doubts have been raised about the
equivalence \cite{Maryland,Richardson}.  

Cutoff regularization is a physically intuitive method for dealing
with divergences in which momenta greater than the scale of new
physics $\Lambda$ are explicitly suppressed.   This can be
implemented by regulating the momentum integrals \cite{BiraThesis} or
by regulating the potential itself~\cite{Lepage}.  
We focus on the latter approach and simplify our discussion by
concentrating on $S$-wave scattering.  We assume the long-distance
physics is properly taken into account in the potential and focus
only on the short-distance EFT potential.  
After defining a
regularized delta-function:  
$\delta_a^3({\bf r}) \equiv e^{-r^2/2a^2}/(2\pi)^{3/2}a^3$, 
the $S$-wave potential 
can be taken as 
\beq
\langle{\bf r'} | V_{\rm CR[G]} | {\bf r}\rangle = 
4\pi a^2 \delta^3({\bf r}-{\bf r'}) 
\left( c\, \delta_a^3({\bf r}) + \frac12 d \, a^2\, \nabla^2
\delta_a^3({\bf r}) + \frac14 e \, a^4\, \nabla^4 \delta_a^3({\bf r}) + \ldots
\right),   
\label{crg1}
\eeq
which is equivalent to including a gaussian suppression of the
transferred momentum ${\bf q}={\bf p}-{\bf p'}$, as seen from the
Fourier transform: 
\be
\langle {\bf p'}| V_{\rm CR[G]} | {\bf p} \rangle \equiv 
V_{\rm CR[G]}({\bf p},{\bf p'}) = 4\pi a^2 \left( c - 
\frac12 d\, {\bf q}^2 a^2 + \frac14 e\; {\bf q}^4 a^4 + \ldots \right) 
e^{-{\bf q}^2 a^2/2} \ .  
\label{crg}
\ee
We will refer to this method as Cutoff Regularization with Gaussian
weighting, or CR[G] for short.  The coefficients $c$, $d$, $\ldots$, are
dimensionless.  The factor of $1/4\pi$ picked up by each additional
term in the Born series requires a $4\pi$ to be factored out in order
to render these coefficients natural \cite{Weinberg}, {\it i.e.\/} of
order unity, as we discuss further in Section~IV.  The coefficients are
determined 
order-by-order from matching to the available data. To fit the first
$n$ constants in the potential requires at least $n$ points of data.
This data should be taken at as low momentum as feasible to minimize
the contribution from terms of ${\cal O}({\bf q}^{2n} a^{2n})$ that
have been omitted. 

We need to make a few comments on the simplicity of
Eqs.~(\ref{crg1}) and (\ref{crg}).   
First, terms that do not contribute to $S$-wave states such as 
${\bf p}\cdot {\bf p'}$ are included in the 
potential (as can be seen by expanding the ${\bf q}^2$ term for
example) to simplify the position-space expression that we use to
evaluate the amplitude \cite{Lepage}.  However, the regularized
delta-function mixes these terms and they end up contributing to $S$-wave
scattering at higher orders in the $a^2$ expansion.  The net effect to
the order we are working is to shift the constant $e$ by a
natural amount.  Since we are only
interested in the naturalness of the constants in this paper and not
their exact value, we will not discuss the effect of such contributions
below.

Second, terms such as $d'\;\nabla \delta_a^3({\bf r}) \cdot\!\nabla\,$
and $\,e'\; a^4 (\nabla^2 \delta_a^3({\bf r}) \nabla^2 + h.c.)$ are
omitted from Eq.~(\ref{crg1}). These terms, when applied with
on-shell solutions of the Lippmann-Schwinger equation, only serve to
modify the coefficients already taken into account in
Eq.~(\ref{crg1}) \cite{FutureWork}. 
Since we fit at most three constants below, we can
take $d'=e'=0$ without any loss of generality to our discussion here. 
Extending the analysis off-shell and to channels other than $S$-wave 
requires a more careful analysis of these contributions.

One advantage to the cutoff method is that an increase of the momentum
cutoff $1/a$ beyond the scale for new physics $\Lambda$  is
signaled 
by unnaturally large coefficients in Eq.~(\ref{crg}) and degraded EFT
estimates for the scattering amplitude.  
This behavior was illustrated by Lepage
\cite{Lepage} and we have verified his results in our analysis.  
It provides a way
to determine $\Lambda$ for an unknown potential or data and suggests
$1/a\sim\Lambda$ as a natural value for the cutoff.
This choice ensures that all the physics
treated correctly at 
low-momentum is still taken into account, but the higher-momentum
physics is suppressed.  Note that while the cutoff is roughly fixed by
the physics, it is not fine-tuned to data as is done for the
coefficients in $V_{\rm CR[G]}$ \cite{Lepage}.  Instead, the cutoff is
a generic parameter of the EFT that dictates the radius of
convergence and naturalness of the constants.

Some disadvantages of the cutoff method are that the nonperturbative
solution of the Lippmann-Schwinger equation requires numerical
techniques and does not transparently exhibit simple power counting.
Furthermore, both chiral and gauge symmetries are broken with a cutoff
and require additional counterterms to restore them.

An alternative approach is to use dimensional regularization, which
preserves the symmetries of the underlying theory and analytically can
be shown to have
simple power counting.  This can be implemented for on-shell
solutions of Eq.~(\ref{LS}) by
using the potential
\beq
V({\bf p},{\bf p'}) = \frac{4\pi}{\Lambda_s^2} \left( c - d\;
\frac{{\bf p}^2+{\bf p'}^2}{2\Lambda_s^2} + e\; 
   \frac{({\bf p}^2+{\bf p'}^2)^2}{4\Lambda_s^4}
+ \ldots \right) \ ,
\label{potential}
\eeq
and using the modified minimal subtraction prescription \drms{}
\cite{KSW1}  or power
divergence subtraction DR[PDS] \cite{KSW2} on the momentum integrals.
Here $\Lambda_s$ is a scale introduced to make the coefficients
dimensionless.
Since \drms{} has no scale (such as a cutoff) associated with the divergences,
the natural size for
$\Lambda_s$ after fitting to the
data
is dictated by physical low-energy scales such as the 
scattering length $a_s$ and effective range $r_e$.  
As shown in Refs.~\cite{KSW1,LukeManohar}, keeping the nonperturbative
amplitude to all orders in $p^2$, 
the momentum expansion is in powers of $p^2 a_s r_e/2$, which
breaks down at very small momentum when the scattering length is large.  
This is a critical issue, since the $NN$ system is
known experimentally to have an almost bound state in the ${}^1S_0$
channel and a weakly bound state in the ${}^3S_1$ channel (the
deuteron).  These are reflected in a large scattering length
corresponding to momentum scales of about $35$ MeV and $90$ MeV respectively.

One way to fix this pathology of \drms{} is to introduce a low-energy
degree of freedom, referred to as the dibaryon, into the lagrangian to
parameterize 
the rapid energy dependence of the amplitude from the
bound or almost bound state~\cite{Kaplan}.%
\footnote{This can also be shown to arise from a certain resummation
of terms in any regularization scheme \protect\cite{vanKolck,Bedaq}.
For \drms{}, however, it is a requirement for a proper radius of
convergence as we will see in Section~IV.}
Taking
the large scattering length explicitly into account removes it from
the residual momentum expansion and improves the radius of convergence.
Another solution is to use DR[PDS], which allows an arbitrary scale
$\mu$ to be included in the subtraction.  This can be shown 
to give a momentum expansion in $p^2 r_e/2(\mu-1/a_s)$, which is well
behaved for large scattering length  if an appropriate $\mu$ 
is chosen \cite{KSW2}. 
With the power counting scheme advocated in \cite{KSW2}, 
DR[PDS] can be shown to produce a $\mu$ independent result.  Since
$\mu=0$ corresponds to \drms{}, this means a specific power counting
scheme cures the problems of dimensional regularization as noted in
\cite{vanKolck,Bedaq}.  For comparison, all references to \drms{}
below refer to the 
naive solution of the Lippmann-Schwinger equation, and DR[PDS] will
refer to this modified power counting scheme. 
All three dimensional regularization techniques can 
be solved analytically for short-range potentials.
  
Finally, we contrast conventional nonrelativistic $NN$ phenomenology to
the effective field theory approach.
The most important difference is that while the phenomenological
potentials incorporate basic pion physics such as one-pion exchange,
none of them has a systematic and complete inclusion of the long-distance
Goldstone boson physics.
But what about the systematic incorporation of short-range physics,
as we  consider here? 
If we take a Reid-style potential \cite{Reid} as an example we can see
similarities. 
The original Reid potential consists of a sum of
Yukawa interactions, with fixed masses chosen as integer multiples of the
pion mass, and coefficients that can be varied to fit the data.
In general for $S$-waves,
\beq
V_{\rm Reid}({\bf p},{\bf p'})= \frac{c_1}{{\bf q}^2+m_1^{\ 2}} +
\frac{c_2}{{\bf q}^2+m_2^{\ 2}} + \ldots \ , 
\label{reidpot}
\eeq
we have a series of Yukawa potentials
that can be viewed as regularized delta functions with different
regulator masses $m_i$.  Since we deal only with the
short range interactions, we are interested in setting these masses on
the order of the scale of underlying physics instead of $m_\pi$.
If the terms in Eq.~(\ref{reidpot}) are combined, 
they generate a Taylor expansion  in ${\bf q}^2$ multiplied
by a function that suppresses large
momentum, similar to the effective field theory potential
Eq.~(\ref{crg}).
However, since each term has a different mass,
a clear separation scale $\Lambda$ is not identified.  

Furthermore, the conventional procedure for fitting the constants has very
different consequences.  
Reid sought the best {\it global\/} fit, varying
the constants to minimize the $\chi^2$ fit to the data over
the entire range of momentum considered.  This means he used a
weighting based only on the errors in the data.  Although this optimizes
the overall agreement with experiment, 
it neglects the fact emphasized in effective
field theories that a theoretical error of order $p^{2n}/\Lambda^{2n}$
is also present in the short-distance physics.

The systematics of an EFT are obscured or lost completely unless this
theoretical error is also taken into account in the weighting of the
fit. 
The ability to relate observables and processes to each other depends
on the ordering of corrections in a well-defined momentum expansion.
Furthermore, only in such a context would constants larger or smaller
than expected be a signal of new physics or symmetry constraints when
matching to QCD.  Therefore, the predictability of the effective field
theory is intimately connected with a consistent treatment of the
error, which requires both the empirical and theoretical errors
to be taken into account.  This is not 
considered in many EFT analyses and so we summarize the
philosophy of effective field theory in the next section and show how
to organize the analysis to reap the full benefits.

\section{Philosophy of EFT Analysis}

\begin{figure}
\begin{center}
\leavevmode
\hbox{
\hspace{-.5cm}
\epsfxsize=3.4in
\epsffile{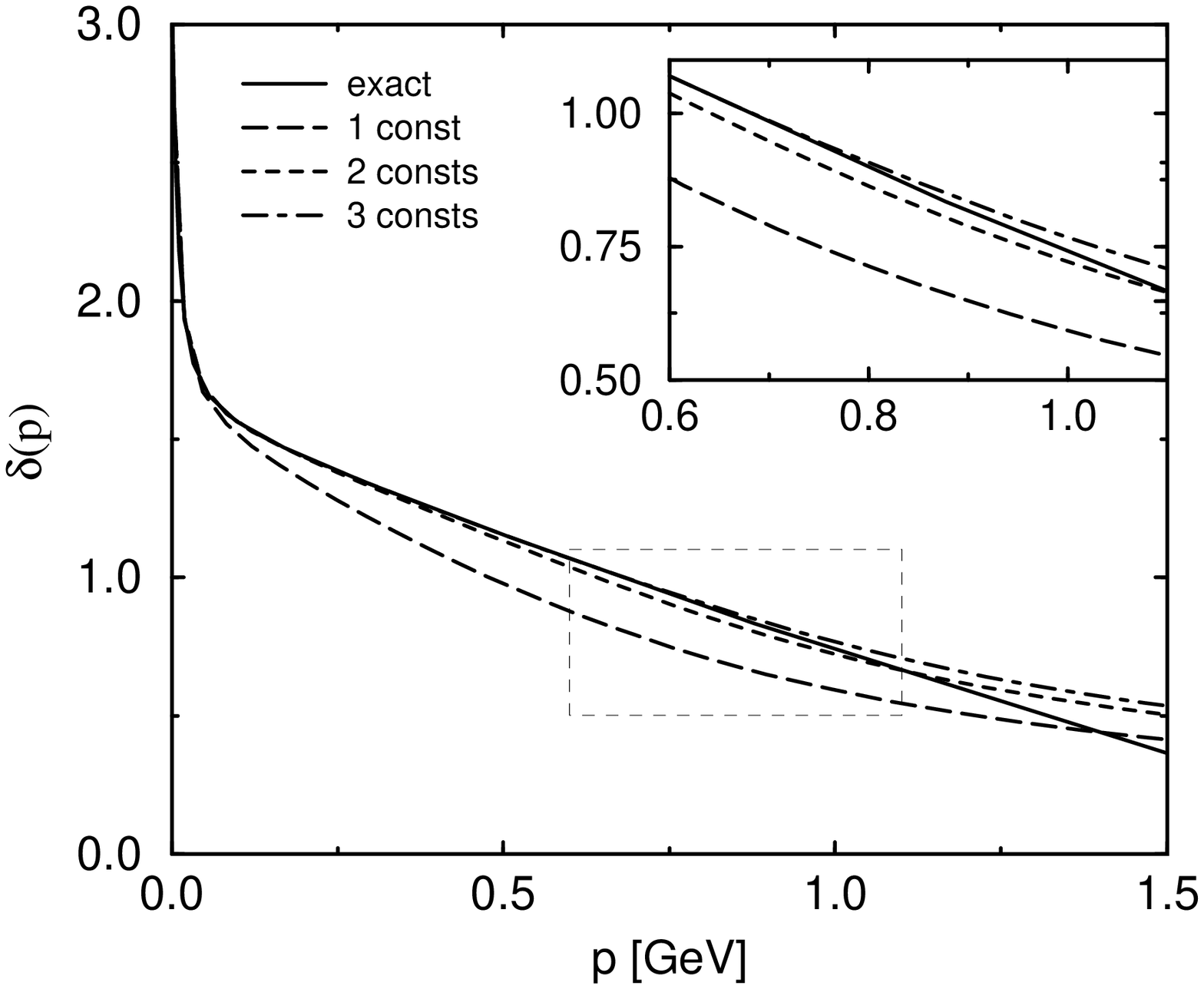}
\hspace{-.4cm}
\epsfxsize=3.4in
\epsffile{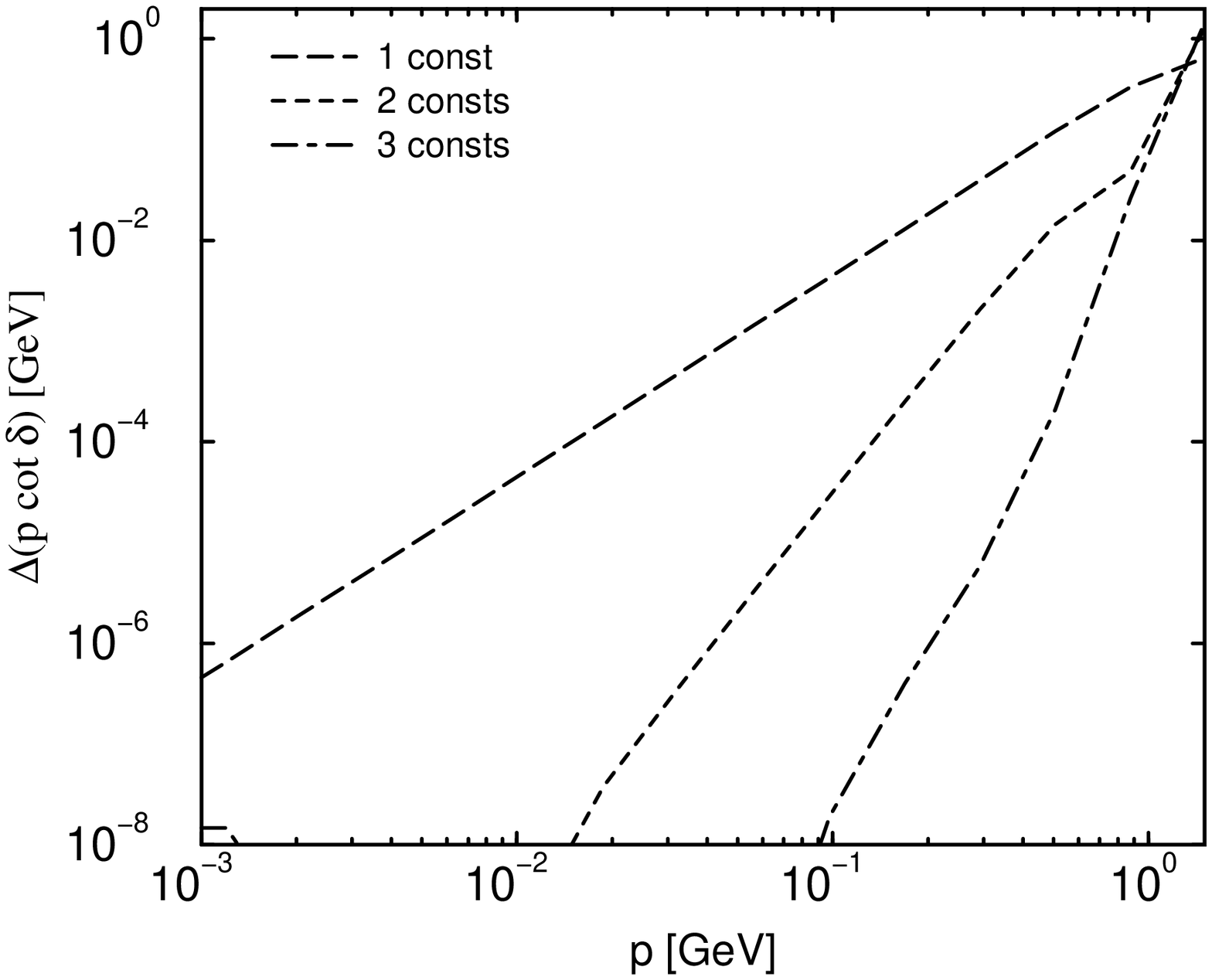}
}
\end{center}
\caption{\label{phaseshift}The phase shift $\delta(p)$ (left) and the
error in $p\cot\delta(p)$ (right), each plotted as a function of
$p$ for the delta-shell potential with a weakly bound state,
as discussed in Section~IV.  The solid line is the exact result
and the dashed lines show the CR[G] fit for one, two, and three
constants.}
\end{figure}

In this section, we describe how to analyze and compare the different
regularization schemes in a manner commensurate with the features of
an effective field theory outlined in the previous
section\footnote{This analysis builds on the comprehensive
discussion in Ref.~\cite{Lepage}.}.  
The potential is not a measurable
quantity, so instead we must use 
a scattering observable such as the phase shift $\delta(p)$ to determine the
constants.  Inserting the potential Eq.~(\ref{potential}) into the
Lippmann-Schwinger equation 
Eq.~(\ref{LS}) and specifying a regularization
scheme, we can solve for the amplitude and determine the
phase shift by one of the following two equivalent relations
\be
T(p)=-\frac{4\pi}{Mp} e^{i\delta(p)} \sin\delta(p) \ ,
\qquad\qquad
-\frac{4\pi}{M} \frac1{T(p)} = p\cot\delta(p) - ip \ .
\label{pcotdel}
\ee
In the left plot of Fig.~\ref{phaseshift}, we show the results from
using CR[G] to 
fit one, two, and three constants in Eq.~(\ref{crg}),
as compared to the exact $S$-wave phase-shift for the delta-shell
potential \cite{Gottfried}, which models the underlying physics and is
discussed in detail in Section~IV. 
A first glance at the plot shows the approximation to the phase shift
improves as more constants in the potential are fit.
However, a second look shows that it is very difficult to gather
any {\it quantitative\/} information from the plot.  At what point the curve
deviates enough to be considered inaccurate is
not clear, and the radius of convergence
of the EFT expansion is completely obscure.  

It is therefore more informative to plot not the phase shift itself,
but the {\em error} in the phase shift: $|\delta_{\rm eff}-\delta_{\rm
true}|$ \cite{Lepage}.  If the effective field theory follows
proper power counting, then this error should improve by two powers of
momentum as each additional coefficient in the potential is fixed [see
Eq.~(\ref{potential})].
However, a simple calculation shows that every term in the momentum
expansion of the phase shift contains the scattering length
\cite{KSW1}. 
This means that the error in the phase shift could be
numerically sensitive to a large scattering length and contaminate the power
counting.

This is not an inherent problem.
We avoid this issue with no loss of generality
by plotting the error of $p\cot\delta(p)$ instead:
\beq
\Delta [ p\cot\delta(p) ] \equiv | p\cot\delta_{\rm eff}(p) -
p\cot\delta_{\rm true}(p) | \ .  \label{eqseven}
\eeq
Since a large scattering length is synonymous
with a near bound state (or pole) in the amplitude, Eq.~(\ref{pcotdel})
shows that this pole is cleanly mapped only to the first constant in a
momentum expansion of $p\cot\delta(p)$.  
It is known from
conventional scattering theory that this combination has a well defined
expansion in $p^2$ for short-range potentials known as the ``effective
range expansion'' 
\beq
p\cot\delta(p) = -\frac1{a_s} + \frac12 r_e p^2 + v_2 p^4 + \ldots \ ,
 \label{ere}
\eeq
which defines the scattering length $a_s$ and effective
range $r_e$.  When long-range potentials are included, 
Eq.~(\ref{ere}) is only valid at  low momentum
or is even inapplicable. 
In this case, one must define a modified effective range expansion
\cite{berger}. 
Since the effective theory contains the same long-distance physics as
the true underlying theory, Eq.~(\ref{eqseven}) can be modified to
have a clean momentum expansion \cite{FutureWork}.
For short range potentials as considered here, it suffices to use
Eq.~(\ref{eqseven}). 

This now gives a rigorous way to fit the coefficients in the effective
potential $V({\bf p},{\bf p'})$ to data.  After evaluating the combination
$p\cot\delta_{\rm eff}(p)$ in the effective theory, we subtract it
from the true result (either data or an exact solution to a model
problem) and fit the difference,
\beq
\Delta p\cot\delta(p) = \alpha + \beta \frac{p^2}{\Lambda^2} + \gamma
\frac{p^4}{\Lambda^4} + \ldots   \ ,
\label{diffER}
\eeq
to a polynomial in
$p^2/\Lambda^2$ to as high an order as possible.  
The convergence rate
of Eq.~(\ref{diffER}) determines the radius of convergence of the
EFT, so we use this observable in our analysis below.  
The use of other observables, such as
the bound-state energy, is discussed in Section~IV. 

Using a spread of
momentum near zero, the polynomial fit should be weighted with both
the expected theoretical error in momentum and any additional
experimental noise.  The resulting coefficients $\alpha, \beta,
\gamma, \ldots$ are then minimized with respect to variations in the
effective potential constants $c,d,e,\ldots$ using an optimization code.
In practice, this method is more robust and numerically stable than
matching the values of $p\cot\delta(p)$ at discrete points to fix the
constants. This allows us to extend the analysis of Lepage
\cite{Lepage} beyond second order as shown in the next section.
We also note that such a procedure is needed when matching to
data even when the EFT observables can be calculated analytically.

The number of coefficients that can be minimized is given by
the number of constants retained in the effective potential.  
We used DPOLFT from package SLATEC \cite{ODE} to find the polynomial fit
and MINF \cite{MINF}, which is based on the Powell method, to carry out the
minimization.  Normal accuracies in minimization using double
precision numbers with this method are $10^{-15}$ or better. 

Plotting the error in $p\cot\delta(p)$ as a function of $p$ on
a log-log plot, we expect a straight line with slope given by the dominant
(lowest) power of $p/\Lambda$ in the error \cite{Lepage}.
As we include more constants, the slope in this error should increase,
signifying the removal of higher powers of $p/\Lambda$.  The second
plot of Fig.~\ref{phaseshift} clearly demonstrates 
the order-by-order improvement
in the amplitude as more constants are added to the effective potential.
With one constant the slope of the error is two [{\it i.e.}, 
${\cal O}(p^2/\Lambda^2)$] and increases by two
with each additional constant. 

This plot also gives a clear 
 graphical interpretation of the radius of
convergence of the EFT
as the point where the lines of error converge.
The second plot of Fig.~\ref{phaseshift} shows that $\Lambda \sim 1\,$GeV.   
For this purpose, the error plot is  much
more informative than plots of the phase shift itself.
Once
the momentum is on the order of the cutoff, the effective theory
breaks down as the short-distance physics is resolved.  Only by
inclusion of  physics at and above the scale $\Lambda$ can the effective
field theory be taken beyond this point.  We will discuss further
consequences of this type of analysis in the next section.  

\section{Illustration with the Delta-shell potential}

In this section, we illustrate the points made above by using EFT
techniques with the different regularization methods to systematically
describe the ``unknown'' short-range physics of a specific example.
We could use $NN$ scattering data, but for our purposes it is more
convenient to use an exactly solvable potential to serve as data in
order to have a clean understanding of what features are important.  
The delta-shell
potential has been used in the past to simulate the large scattering
length found in $NN$ scattering \cite{Gottfried}.  
Kaplan used
this potential to illustrate the benefits of the dibaryon approach
\cite{Kaplan}.  He found upon considering $NN$ scattering that the
inclusion of long-distance pion physics did not change the
conclusions.  This agrees with the
experience of other authors \cite{Maryland,KSW1} that the addition
of pions as long-range interactions does not affect the properties
of the short-range expansion.
The delta-shell potential
is therefore a sufficient model for our purposes here.

The potential can be written in terms of the nucleon mass $M$, the
coupling $g$, and the range of the potential $r_0$.  This short-range
potential represents the new physics of our underlying theory, and so we
take $r_0=1/\Lambda$ below,
\beq
V_{\rm true}(r) = -g \frac{\Lambda}{M}\,
\delta\left(r-\frac1\Lambda\right) .  \label{thispotential}
\eeq
It has exactly one bound state for $g\ge1$ and no bound states
for $g<1$.  Scattering with $p >
\Lambda$ probes the details of the potential, so we 
expect $\Lambda$ to be the radius of convergence of a well-tuned EFT.
The scattering length becomes very large for $g$ near 1,
whereas the effective range (and the rest of the terms in the
effective range expansion) are of natural size for all $g$:
\be
a_s = \frac{g}{g-1}\; \frac1{\Lambda}\ ,
\qquad\qquad
r_e =  \frac{2(g+1)}{3g} \; \frac1{\Lambda}\  .
\label{scat}
\ee
The scattering length in 
the ${}^1S_0$ channel of $NN$-scattering can be
modeled by choosing
$(g,\Lambda)=(0.99,m_\rho)$.  
This potential Eq.~(\ref{thispotential}) with different $g$'s will be the
``laboratory''  from which we compare the different regularization
schemes.  Results for actual measured data and the inclusion of 
pions will be discussed in a future publication \cite{FutureWork}.

We will first gain some intuition by graphically reproducing Kaplan's
result that \drms{} has a small radius of convergence if the
scattering length is large. From Eq.~(\ref{scat}), we
see that choosing $g=0.99$ gives a scattering length one
hundred times
larger than choosing $g=-10$.  At the same time, we investigate the
effect from the presence of a bound state by taking $g=1.01$. 
The effective potential is given by
Eq.~(\ref{potential}) with the mass scale $\Lambda_s$ associated
with the inverse delta-shell radius and the prescription of using \drms{} on
all divergent integrals.

\begin{figure}
\begin{center}
\leavevmode
\hbox{
\hspace{-.5cm}
\epsfxsize=3.4in
\epsffile{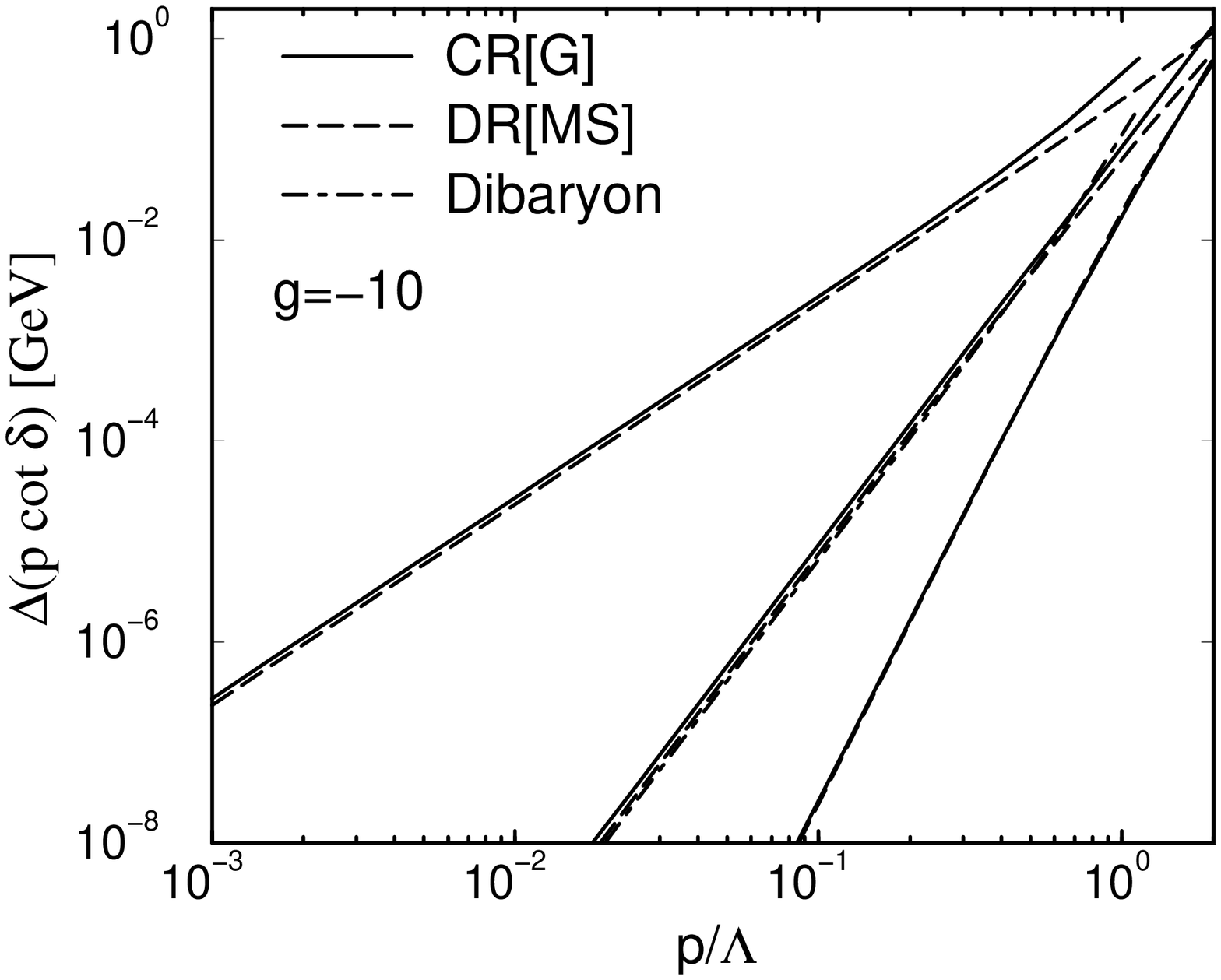}
\hspace{-.4cm}
\epsfxsize=3.4in
\epsffile{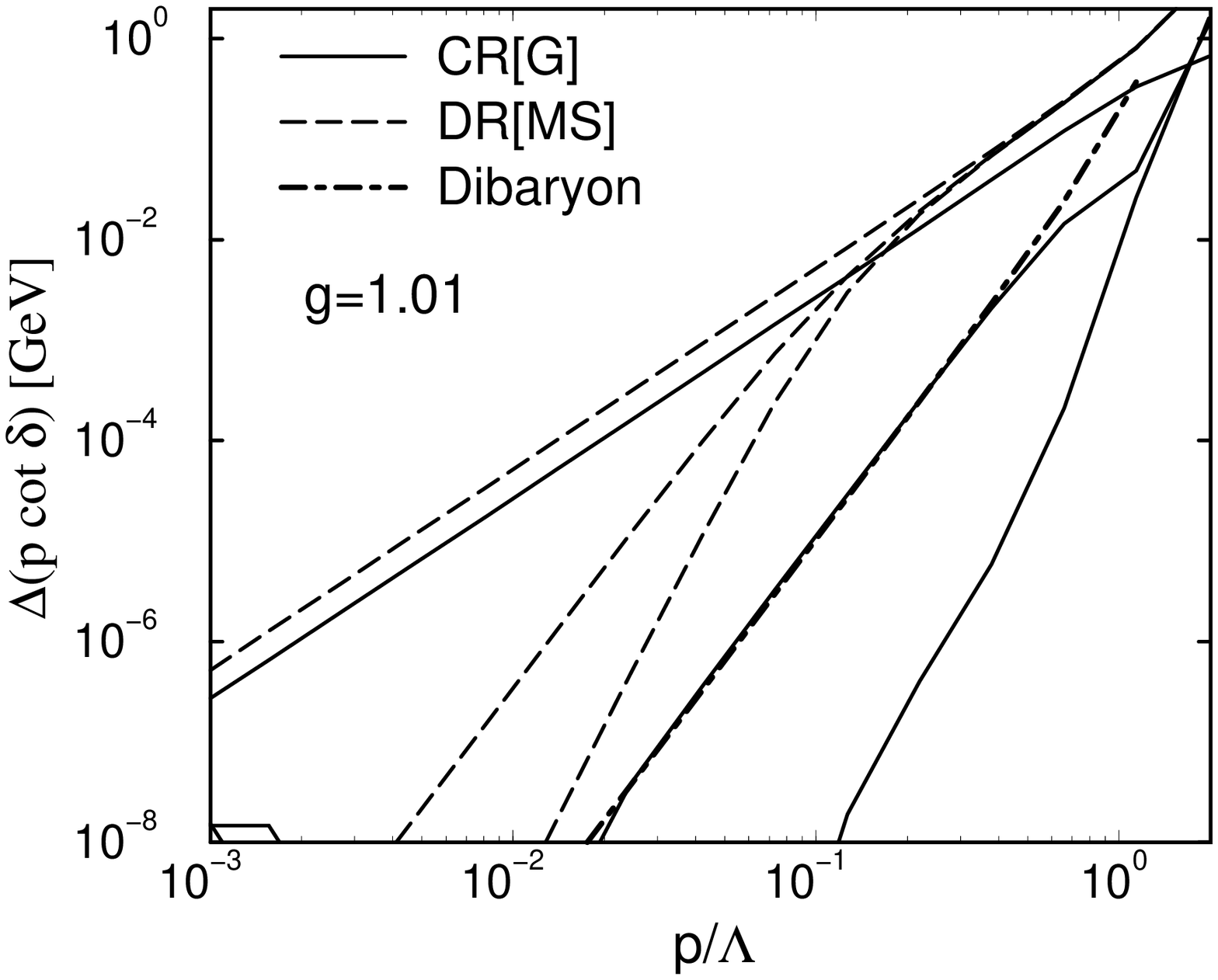}
}
\end{center}
\caption{\label{err1}The error in $p\cot\delta(p)$ plotted as a function of
$p/\Lambda$ for a small scattering length without a bound state $g=-10$
and for a large scattering length with a bound state $g=1.01$.}
\end{figure}

As mentioned above, the momentum expansion for \drms{} can be shown
analytically to be $p^2 a_s r_e/2$ \cite{KSW1,Kaplan}.  This implies
using Eq.~(\ref{scat}) that the radius of convergence for $g=1.01$ and
$0.99$
should be roughly $1/10$ that of the $g=-10$ case.  Fixing the
constants in the potential Eq.~(\ref{potential}) by matching
$p\cot\delta(p)$ as outlined in the previous section, we produce the
results in Fig.~\ref{err1}.  The dashed 
lines show the \drms{} results for one, two, and three constants
respectively.  Indeed all three lines converge to $p/\Lambda\sim1$ for
$g=-10$ and $p/\Lambda\sim0.1$ for $g=1.01$.  The results for $g=0.99$
fall on top of the $g=1.01$ results and are therefore not shown.  This
implies the presence of a bound state as opposed to an almost bound
state does not matter, but the size of
the scattering length does.  
It would be difficult to
draw these conclusions had we only plotted the phase shift itself
(the left plot of Fig.~\ref{phaseshift}).  Of course, we can also show
these results
analytically for this simple model, 
but the analysis applies much more generally, when part or all of the
calculation is done numerically.

The constants $c$, $d$, and $e$ are given in Tables~\ref{tab2} and
\ref{tab3}.  Their
values are calculated numerically to at least 8 digits to produce the
accuracy of Fig.~\ref{err1} and agree with the analytical values.  This
serves as a check on our numerical routines.  The 
correlation between the naturalness of the constants and the radius of
convergence is apparent.  
The constants for the $g=1.01$ case are extremely large,
reflecting the breakdown of the effective field theory for \drms{} much
below the expected scale $\Lambda$.

\begin{table}[t] 
  \caption{\label{tab2}Effective potential for $g=-10$ (small scattering
    length) to three different
    orders for different regularization schemes.  $\Lambda a=1$ for CR[G] 
     and $\mu=\Lambda$ for DR[PDS].} 
\vspace{.5cm}
\begin{tabular}{c|ccc|ccc|ccc}
 & \multicolumn{3}{c|}{\drms{}} & \multicolumn{3}{c|}{DR[PDS]} &
        \multicolumn{3}{c}{CR[G]} \\  
   & $c$ & $d$ & $e$ & $c$ & $d$ & $e$ & $c$ & $d$ & $e$ \\  \hline
${\cal O}(p^2/\Lambda^2)$ & 0.758 & --- & --- &
                            8.33  & --- & --- &
                            1.49  & --- & ---  \\
${\cal O}(p^4/\Lambda^4)$ & 0.758 & $-0.206$ & --- &
                            8.33  & $-25.0$  & --- &
                            2.67  & $-1.04$ & ---  \\
${\cal O}(p^6/\Lambda^6)$ & 0.758 & $-0.206$ & 0.0672\ &
                            8.33  & $-25.0$  & 38.2\ &
                            2.67  & $-1.58$ & 3.75\  \\
\end{tabular}
\vspace*{1.0cm}

  \caption{\label{tab3}Effective potential for $g=1.01$ (large scattering
    length) to three different
    orders for different regularization schemes.  $\Lambda a=1$ for CR[G] 
    and $\mu=\Lambda$ for DR[PDS].} 
\vspace{.5cm}
\begin{tabular}{c|ccc|ccc|ccc}
 & \multicolumn{3}{c|}{\drms{}} & \multicolumn{3}{c|}{DR[PDS]} &
        \multicolumn{3}{c}{CR[G]} \\  
   & $c$ & $d$ & $e$ & $c$ & $d$ & $e$ & $c$ & $d$ & $e$ \\  \hline
${\cal O}(p^2/\Lambda^2)$ & 84.2 & --- & --- &
                            $-0.842$  & --- & --- &
                            $-1.42$  & --- & ---  \\
${\cal O}(p^4/\Lambda^4)$ & 84.2      & $-5.64\times 10^3$ & --- &
                            $-0.842$  & $-0.564$ & --- &
                            $-0.946$   & $0.842$ & ---  \\
${\cal O}(p^6/\Lambda^6)$ & 84.2      & $-5.64\times 10^3$ 
                                         & $3.78\times 10^5$\ &
                            $-0.842$  & $-0.564$           & $-0.152$\ &
                            $-0.937$   & $0.618$           & $-0.181$\  \\
\end{tabular}
\end{table}

One way to fix this behavior is to introduce the dibaryon
\cite{LukeManohar,Kaplan,Bedaq}. This takes the large scattering length into
account by explicitly introducing a low-energy $s$-channel degree of
freedom into the effective lagrangian.  For $g>0$ or $g<-1$, 
the potential can be written as
\be
V_{\rm dibaryon}({\bf p},{\bf p'}) = C - \frac{y^2}{E+\Delta};
\qquad
C = \frac{2\pi}{M\Lambda},
\mbox{\ \ }
y^2 = \frac{3\pi\Lambda}{M^2} \frac{1+g}{g},
\mbox{\ \ }
\Delta = \frac{3\Lambda^2}{2M} \frac{1-g}{g},
\label{dibar}
\ee 
with $E=p^2/M$ always kept on-shell.  Since there seem to be three
constants fit in Eq.~(\ref{dibar}), one might think the dibaryon will
have an error of ${\cal O}(p^6/\Lambda^6)$.  However, the dibaryon
amplitude is only matched to second order in the momentum when
deriving the relations in 
Eq.~(\ref{dibar}) \cite{Kaplan} and indeed shows an error of ${\cal
O}(p^4/\Lambda^4)$ for the two values of $g$ in Fig.~\ref{err1}
(plotted as the dot-dashed line).  The  
slope and magnitude of the error do not depend on the scattering
length, as expected.

We next repeat the calculations using the cutoff regularization method
CR[G] with $1/a=\Lambda$.  There are various ways to numerically solve
the Schr\"odinger equation with a cutoff, but
we have found the following procedure to be particularly efficient
and numerically robust.  
First, the variable phase method is used to
solve for the phase shift.
This is a differential equation,
\beq
\delta'(r) = - \frac{M}{p}\; V(r)\; \sin^2\!\left( pr + \delta(r)
\right)
\qquad\mbox{with\ } \delta(0)=0 \ ,
\eeq
which expresses the change in the phase shift as
the potential is built up from zero at $r=0$ to its full value at
$r=\infty$.  
The boundary condition ensures that the full phase shift given by
$\delta(\infty)$ is zero in 
the absence of a potential and defines the otherwise ambiguous
multiple of $\pi$ in the phase shift. We use the routine ODE from
package ODE \cite{ODE} to solve the differential equation.  To obtain
the accuracy shown in the plots,
we needed to do a
weighted polynomial fit of $\Delta p\cot\delta(p)$ up to
$p/\Lambda=0.1$.

The solid lines in Fig.~\ref{err1} show that CR[G] does work
regardless of the scattering length.  In fact, 
with $\Lambda \sim 1/a$, the result is just as
good as the dibaryon for the same number of constants.  The values for
the constants are given in Tables~\ref{tab2} and \ref{tab3}, showing
they are all 
natural for both $g$'s (although the third constant is somewhat 
small for $g=1.01$).

Note that as more constants are fixed, the lower-order constants are
modified.  
This occurs because even after truncating the potential in
Eq.~(\ref{crg}) to a given order, it still contains all orders in
$p^2$ from the gaussian factor.
The nonperturbative solution of the Lippmann-Schwinger equation
therefore can generate terms of {\em any} order in $p^2/\Lambda^2$.
The amplitude itself is matched to the true result order-by-order in
the momentum so the power counting of the potential is destroyed.
This consequence of the cutoff regularization is not necessarily
relevant since
the potential is not an observable. 
It is interesting to note, however, that these
modifications are relatively small. This is also true when adding the long
distance physics of the pion in fitting to actual $NN$ data
\cite{FutureWork}.

In summary, Fig.~\ref{err1} shows that all regularization
schemes considered so far  produce useful effective field theories for
a small scattering length, but \drms{} fails for large scattering
length.  A failure of the power counting in powers of $p/\Lambda$ is
reflected in unnatural constants in the potential.

\begin{figure}
\begin{center}
\leavevmode
\hbox{
\hspace{-.5cm}
\epsfxsize=3.4in
\epsffile{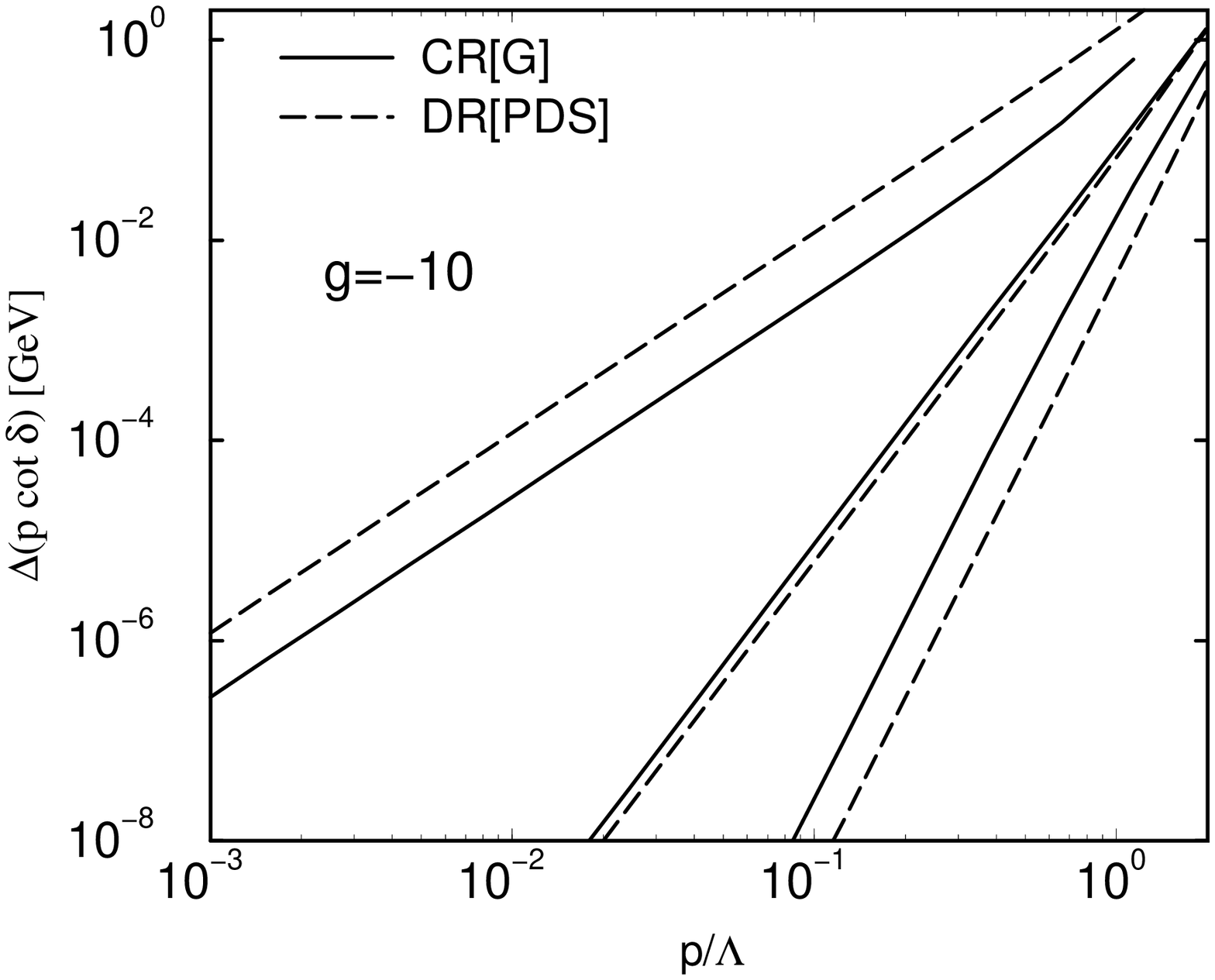}
\hspace{-.4cm}
\epsfxsize=3.4in
\epsffile{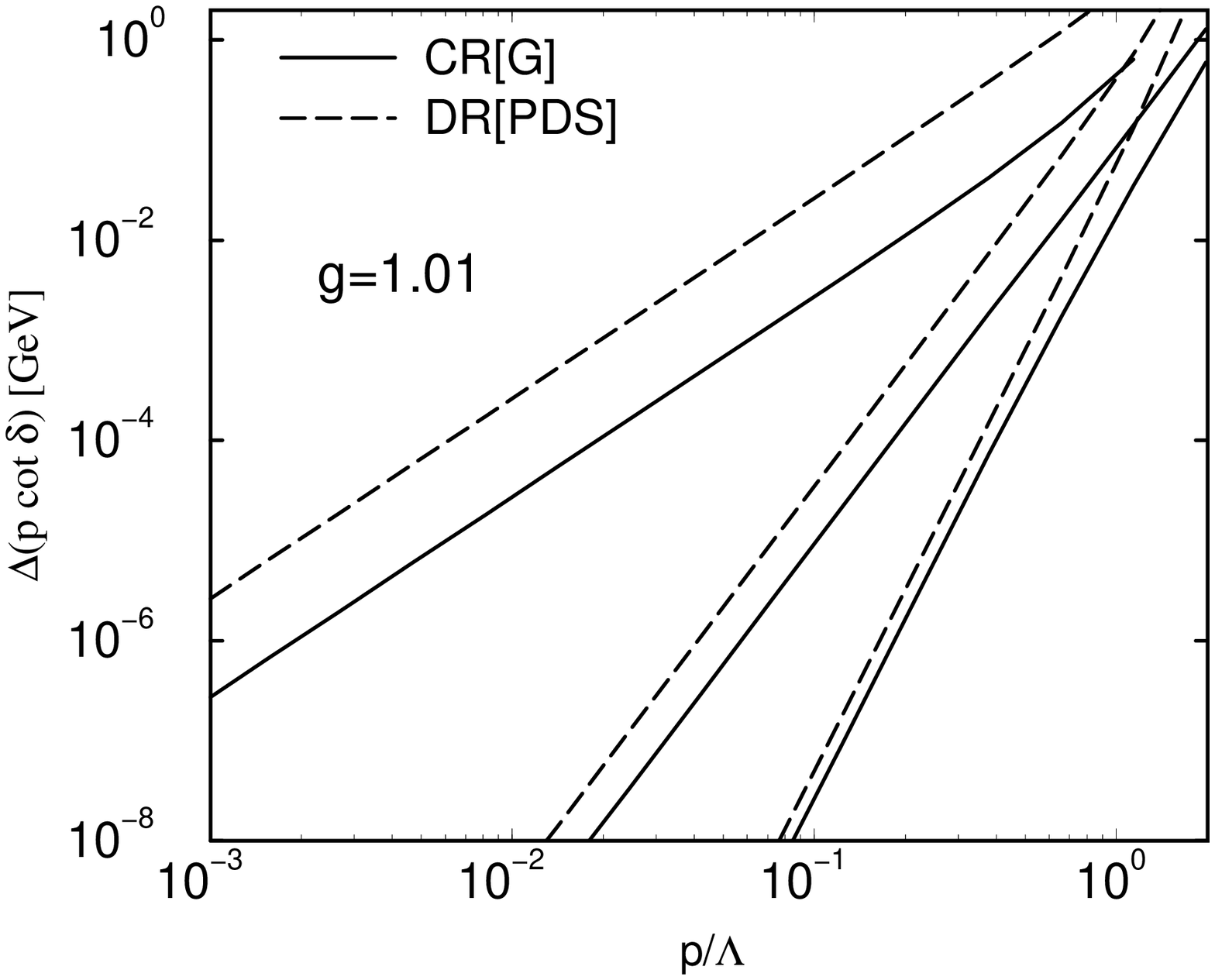}
}
\end{center}
\caption{\label{err2}The error in $p\cot\delta(p)$ plotted as a function of
$p/\Lambda$ for a small scattering length without a bound state $g=-10$
and for a large scattering length with a bound state $g=1.01$.}
\end{figure}

We now focus on the most recently proposed regularization scheme,
DR[PDS] \cite{KSW2}.  
The potential is given by Eq.~(\ref{potential}), but with a
further subtraction of an arbitrary mass scale $\mu/4\pi$
from the linearly divergent integrals.  This term  mimics the
behavior of the cutoff method, although  still allowing a simple
enough form to be
solved analytically.  
An additional prescription compared to the \drms{} case is an
expansion of observables to the same order in $p^2$ as the potential
Eq.~(\ref{potential}). 
If this is not done, the results are $\mu$ dependent, with
$\mu=0$ reproducing \drms{} and $\mu$
larger than the nucleon mass approaching the 
CR[G] result in Fig.~\ref{err1}. 

Since we are only dealing with a short-range potential, the DR[PDS]
prescription reproduces the effective range expansion Eq.~(\ref{ere})
by construction. 
The DR[PDS] results in Fig.~\ref{err2}  
are therefore $\mu$ independent, although the constants still depend
on $\mu$. We take $\mu=\Lambda$ to compare with CR[G]. 
 This produces natural constants for
$g=1.01$ but somewhat unnatural ones for $g=-10$ as seen in
Table~\ref{tab2}.  This is an accidental consequence of the momentum
expansion being in powers of $p^2 r_e/2(\mu-1/a_s)$, so that for $g=-10$
and $\mu=\Lambda$ [see Eq.~(\ref{scat})] the
denominator is nearly zero.  The best result in this case occurs for
$\mu=0$, which reproduces the \drms{} result and has natural constants.
This implies that the scale $\mu$ is not 
functionally equivalent to the cutoff
$1/a$ in CR[G] since it does not always signal the onset of new
physics at the scale $\Lambda$.  However, the $\mu$ independence
of the scattering length shows this prescription for power counting
produces a satisfatory radius 
of convergence, even for \drms{} ($\mu=0$) \cite{vanKolck,Bedaq,KSW2}. 

For both large and small scattering length, DR[PDS] does quite
well, with a radius of convergence $p/\Lambda\sim 1$.  The CR[G]
result is better for one constant since the cutoff generates
an effective range $r_e$ close to the true result.  Overall,
DR[PDS] is a convenient method to produce reasonable
analytical results, and depending on the problem at hand either CR[G]
or DR[PDS] may be suitable.  
One should note, however, that only DR[PDS] provides a strict diagram
by diagram power counting \cite{KSW2}.

Finally, we return to
the Reid potential \cite{Reid}.   
The original Reid analysis used a
global fit and only one adjustable parameter in each channel,
for a result with
approximately the same error (roughly a few percent of the data) at
all momenta.
However, we can apply the EFT fitting procedure instead.  
If we  use Yukawa
masses comparable to $\Lambda$, we anticipate similar results
to CR[G].
Indeed, if a low-momentum fit is done to the constant
$c_1$, the error plot is
similar to the CR[G] result with one constant (Fig.~\ref{err1}).
Adding a second short-range Yukawa does as well as CR[G] with two
constants, since the Yukawas play off each other to allow the next order
error in ${\bf q}^2$ to be removed.
This interplay becomes increasingly complex at higher orders.
Furthermore,
the additional mass scales  obscure (or smear out) 
the role of
$\Lambda$ as a  scale that separates the known from the
unknown physics in effective field theories.
Traditional $NN$ potentials such as Reid 
are well suited for global fits.
Systematic predictions with controlled error estimates are more
properly analyzed using
an effective field theory.


We now turn to an investigation of the binding energy.  If the EFT is
truly reproducing the S-matrix of the underlying theory order-by-order
in a momentum expansion, it should reproduce the binding energies and
other observables to the same order of accuracy as the phase shifts.
We therefore use the binding energy prediction as a consistency check
for our candidate effective field theories.

We have already fit the potentials to a given order by the scattering
phase shifts above, and we use these potentials 
without adjustment to solve for the
binding energy. This can be done analytically for the DR schemes by
finding the poles in the scattering amplitude.
The exact binding energy $E_{\rm bind}$
for the delta-shell potential is given by solving the equation
\beq
\frac1{g} = \frac{1-e^{-2\eta}}{2\eta},
\qquad\qquad
\eta=\frac{\sqrt{M E_{\rm bind}}}{\Lambda} .
\eeq
There is only one bound state to predict in the delta-shell
potential, and if it is shallow enough, even the effective range
expansion with the values of $a_s$ and $r_e$ can determine its value.
A better test is to increase $g$ until $E_{\rm bind}$ is large and
on the order of $\Lambda$, and use the EFT to determine the accuracy of the
binding energy prediction as a function of this variation.  
If a true radius of convergence is present, the effective field theory
should break down for $E_{\rm bind}/\Lambda\sim 1$.  The binding energy
is $6.48\times 10^{-2}$ MeV for $g=1.01$ but quickly increases
to $812$ MeV for $g=2.5$. 

\begin{figure}
\begin{center}
\leavevmode
\hbox{
\hspace{-.5cm}
\epsfxsize=3.4in
\epsffile{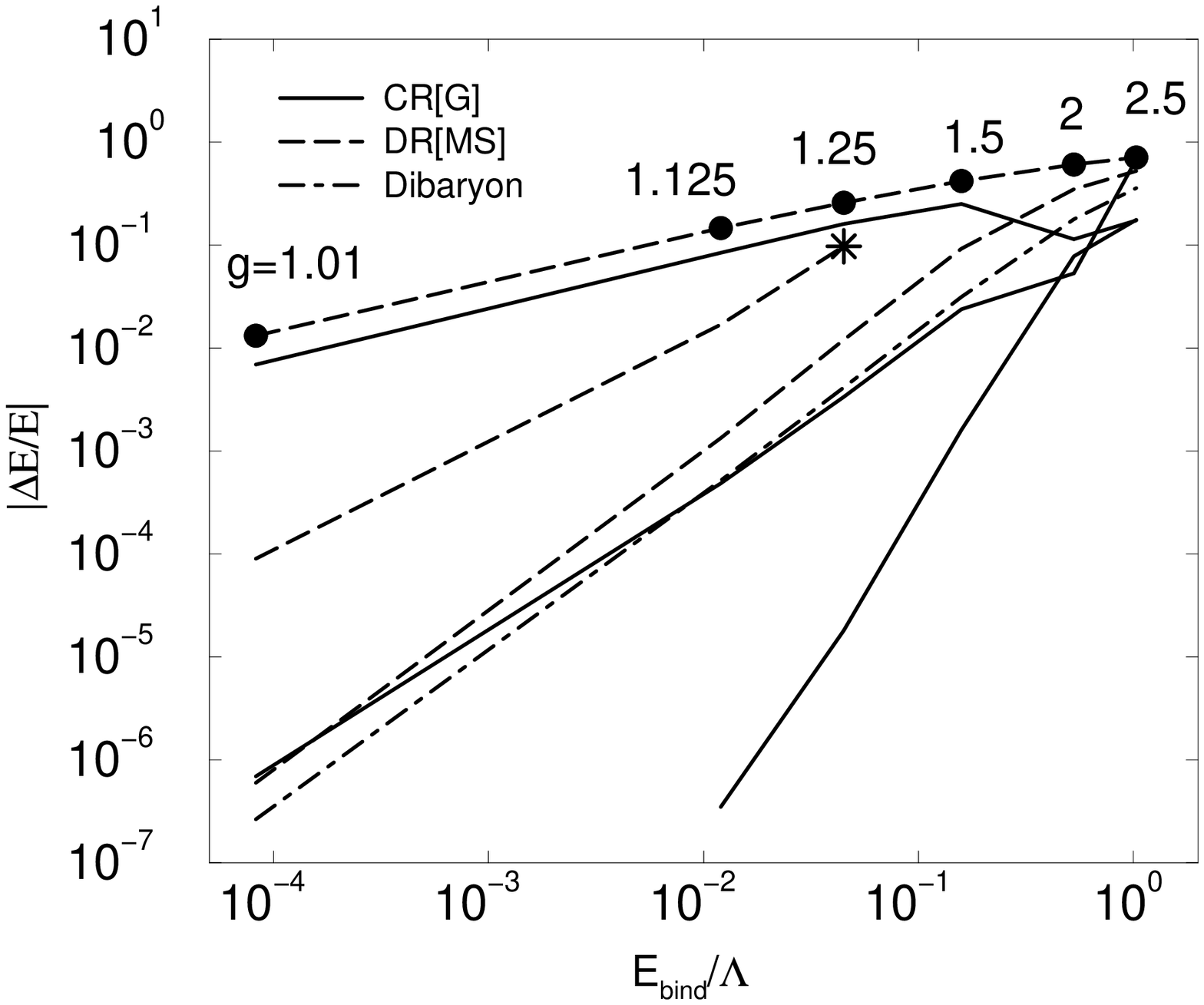}
\hspace{-.4cm}
\epsfxsize=3.4in
\epsffile{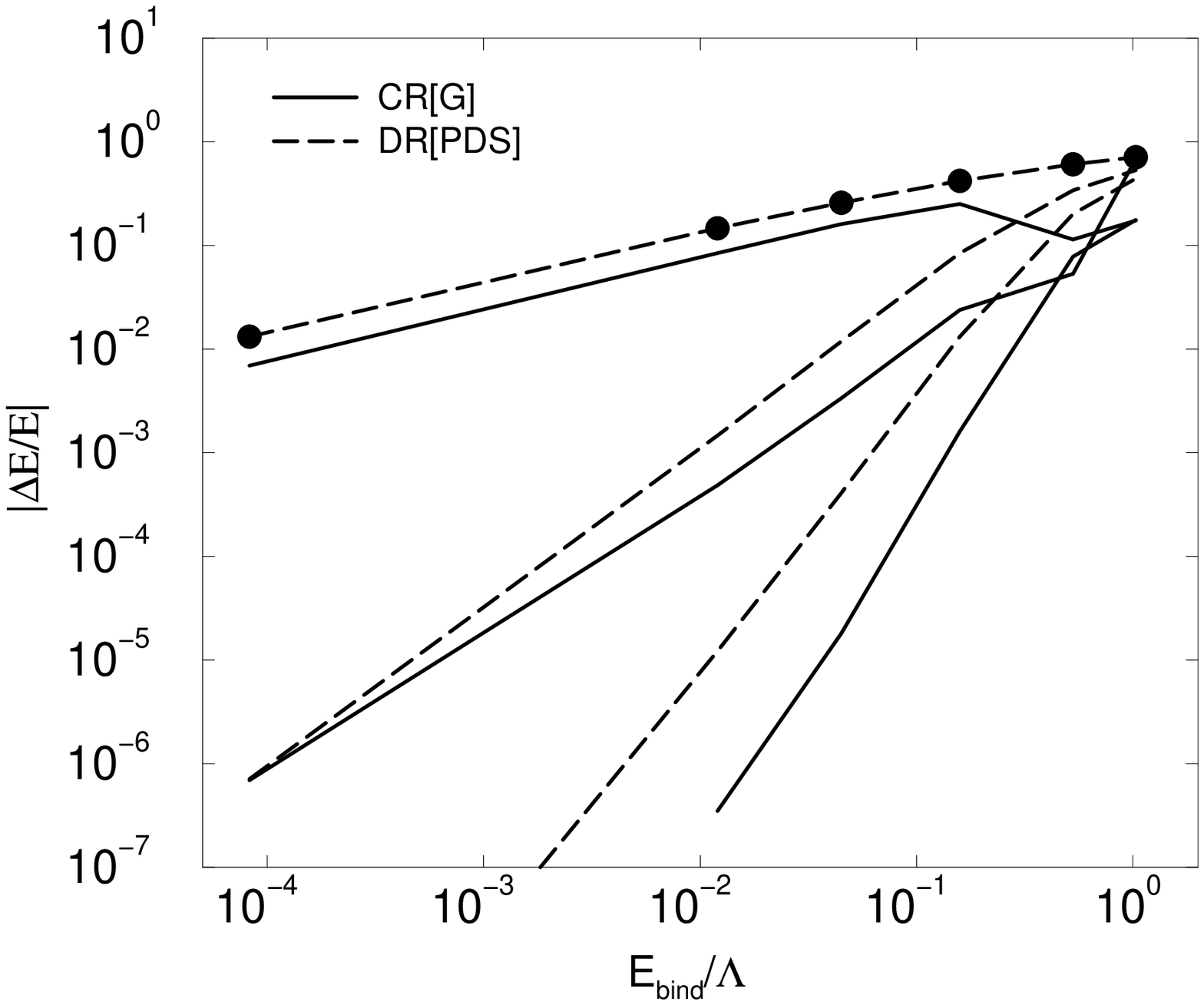}
}
\end{center}
\caption{\label{bind}The error in the binding energy for the candidate
effective field theories for representative values of $g$ from $1.01$
to $2.5$.  
The star signifies the absence of real binding energies for 
\protect\drms{} with two
constants at the values of $g>1.25$ considered.}
\end{figure}

We plot the relative error in the binding energy in Fig.~\ref{bind}.
Both CR[G] and DR[PDS] show the clear power counting behavior and
proper radius of convergence expected from a true effective field
theory.  This gives a graphical verification that the errors in the
binding energy really do follow power counting rules.
We have checked that the same behavior is seen when plotted as a
function of the average momentum $\sqrt{\langle p^2\rangle}/\Lambda$.
The dibaryon result also follows the expected error scaling.
In contrast, the deficiencies of \drms{} regularization
seen for the phase shifts
are manifested here as binding energies that do not follow the EFT
error scaling, improving to a lesser degree than expected.  
In addition, for values of $g > 1.25$ with two
constants, the \drms{} S-matrix shows no bound state with
a real energy.

Therefore, our results show that most 
regularization procedures demonstrate
the characteristics of a systematic predictive effective
field theory.  The fit of more and more constants in the effective
potential improves the predictive power order-by-order in the momentum
expansion.  The radius of convergence of the EFT is independent of the
scattering 
length and is given by the scale where new physics enters.

\section{Conclusions}

Attempts to apply effective field theory  methods to the
nonrelativistic nucleon-nucleon  scattering problem and then to
the nuclear many-body problem have been stalled because of
controversies concerning the nature and limitations of an EFT
expansion when used nonperturbatively.  The familiar dictum that
``calculated observables are independent of the regularization
method'' has been questioned in this context.  To help resolve these
issues we have made a direct comparison of the various regularization
approaches.

We have applied the error analysis suggested by Lepage \cite{Lepage}
to a model problem to compare cutoff regularization, two forms of 
dimensional regularization, and the dibaryon approach in the context
of nonperturbative, nonrelativistic effective field theories.
This analysis focuses on a key signature of EFT behavior: the
systematic scaling of errors with momentum or energy.
We summarize some points  made by Lepage about applying cutoff
effective field theory \cite{Lepage}:
\begin{itemize}
  \item The cutoff potential is not an observable
   and is not amenable to power counting.  
   Individual constants can change as higher orders are taken into
   account, but predictions for observables are still systematically improved. 
  \item Fits to data should be weighted by both the uncertainty in the
  data and the expected theoretical error from power counting. This
  applies to any regularization scheme.
  \item The cutoff should not be taken to infinity but
  only roughly adjusted to minimize the error,
  which identifies the resolution scale of the underlying short-distance
  physics. Using the cutoff as a fine-tuned parameter is not as
  effective as the introduction of a new low-energy constant in the
  potential. 
\end{itemize}
Our results verify these points.   We also observed that natural
coefficients are correlated with an optimal radius of convergence.

New numerical procedures set forth in this paper
allow us to work to third order and beyond
in the EFT expansion, which is necessary to obtain a clear
graphical determination of the radius of convergence for a given
observable.
Such an analysis is required for a systematic fit to data regardless
of the regularization scheme. 

We find that all of the regularization methods except for
dimensional regularization with modified minimal subtraction
are consistent with basic features expected from a useful effective
field theory:
\begin{itemize}
  \item Each additional order in the potential leads to a systematic
  improvement in the amplitude.
  \item The radius of convergence for this improvement, when
  optimized, is dictated by
  the scale of new physics.
  \item Other observables are predicted with the same accuracy as the
  amplitude at each successive improvement.
\end{itemize}
Our results are consistent with the analysis of van Kolck that, with
proper resummations, any effective field theory for short-range
interactions is equivalent to an effective range expansion to the same
order \cite{vanKolck}.

The CR[G] and DR[PDS] regularization schemes are each suitable 
for developing effective field theories of many-nucleon systems.
In future work we will use both schemes
in extending our fitting procedure and error analysis
to $NN$-scattering (including pions and other channels)
and then to nuclear matter.

\acknowledgments

We thank T.~D.\ Cohen, D.~B.\ Kaplan, G.~P.\ Lepage, and R.~J.\ Perry
for useful discussions. 
This work was supported in part by the 
National Science Foundation
under Grants No.\ PHY--9511923 and PHY--9258270.

\end{document}